\documentclass[journal,transmag]{IEEEtran}
\usepackage{ulem}
\usepackage{cite}

\ifCLASSINFOpdf
\usepackage[pdftex]{graphicx}
\else
\fi
\usepackage{color}
\usepackage{booktabs}
\usepackage{multirow}

\usepackage{amsmath}

\usepackage{ulem}

\begin{document}

%
\title{Computational search for ultrasmall and fast skyrmions in the Inverse Heusler family}


\author{\IEEEauthorblockN{Yunkun Xie\IEEEauthorrefmark{1},
Jianhua Ma\IEEEauthorrefmark{1},
Hamed Vakilitaleghani\IEEEauthorrefmark{2}, 
Yaohua Tan\IEEEauthorrefmark{3}, and
Avik W. Ghosh\IEEEauthorrefmark{1,2}}
\IEEEauthorblockA{\IEEEauthorrefmark{1}School of Electrical and Computer Engineering,
University of Virginia, Charlottesville, VA 22903 USA}
\IEEEauthorblockA{\IEEEauthorrefmark{2}Department of Physics,
University of Virginia, Charlottesville, VA 22903 USA}
\IEEEauthorblockA{\IEEEauthorrefmark{3}Synopsys inc. CA 94043 USA}
\thanks{Corresponding author: Yunkun Xie (email: yx3ga@virginia.edu).}}

\markboth{Journal of \LaTeX\ Class Files,~Vol.~14, No.~8, August~2015}%
{Shell \MakeLowercase{\textit{et al.}}: Bare Demo of IEEEtran.cls for IEEE Transactions on Magnetics Journals}
%



\IEEEtitleabstractindextext{%
\begin{abstract}
Skyrmions are magnetic excitations that are potentially ultrasmall and topologically protected, making them interesting for high-density all-electronic ultrafast storage applications. While recent experiments have confirmed the existence of various types of skyrmions, their typical sizes are much larger than traditional domain walls,  except at very low temperature. In this work, we explore the optimal material parameters for hosting ultra-small, fast, and room temperature stable skyrmions. As concrete examples, we explore potential candidates from the inverse Heusler family. Using first-principles calculations of structural and magnetic properties, we identify several promising ferrimagnetic inverse Heusler half-metal/near half-metals and analyze their phase space for size and metastability.
\end{abstract}

}

\maketitle

\IEEEdisplaynontitleabstractindextext

%
\IEEEpeerreviewmaketitle

\section{Introduction}
%
%
%
%
\IEEEPARstart{M}{agnetic} skyrmions have topological spin textures that are stabilized by their antisymmetric exchange or Dzyaloshinskii-Moriya interaction (DMI). Their vortex-like spin configurations were predicted to exist stably at the
nanometer scale in bulk non-centrosymmetric materials as well as thin film heterostructures\cite{bogdanov1989thermodynamically,bogdanov2001chiral}. Substantial progress has been made in observing skyrmions and skyrmion lattices in multiple systems by means of neutron scattering, Lorentz transmision electron microscopy,  scanning tunneling microscopy and X-ray holography. Reported material systems supporting skyrmions include the B20 family (such as MnSi\cite{neubauer2009topological,pappas2009chiral}, FeGe\cite{huang2012extended}, $\mathrm{Fe_{0.5}Co_{0.5}Si}$\cite{yu2010real}), multiferroic materials\cite{seki2012observation}, tetragonal inverse heuslers\cite{nayak2017magnetic}, thin film Fe/Ir\cite{heinze2011spontaneous,romming2013writing} FeCoB\cite{woo2016observation} and amorphous ferrimagnets such as $\mathrm{Gd_{44}Co_{56}}$\cite{caretta2018fast}. These exciting results bring up possible applications of skyrmions in reliable high density all electronic information processing and storage, such as racetrack memory, where information is stored in magnetic domains and driven by a current, as has been demonstrated in a nanowire \cite{parkin2015memory}.

One of the challenges with racetrack memory is the pinning of domain walls at defect sites. Skyrmions can be manipulated using spin transfer torque or spin-orbit torque much like domains, with the added advantages of smaller size and some amount of topological protection, reducing the threshold current for activation around defects\cite{sampaio2013nucleation}. However, the reported skyrmions are either too big ($\sim100\,\mathrm{nm}$) or exist only at low temperatures\cite{romming2013writing}. There are also issues specific to skyrmion dynamics such as the skyrmion Hall effect, where the Magnus force swivels the skyrmions away from a linear trajectory towards the device edges for potential annihilation. In this paper, we address some of these issues from a materials point of view. We provide simple equations and phase space for specific material suggestions through first-principles and micromagnetic simulations.

\section{Towards small\&fast skyrmions}
\subsection{Stable isolated skyrmion}
\label{sec:eq_skm}
\begin{figure}[ht]
\centering
\includegraphics[width=3.5in]{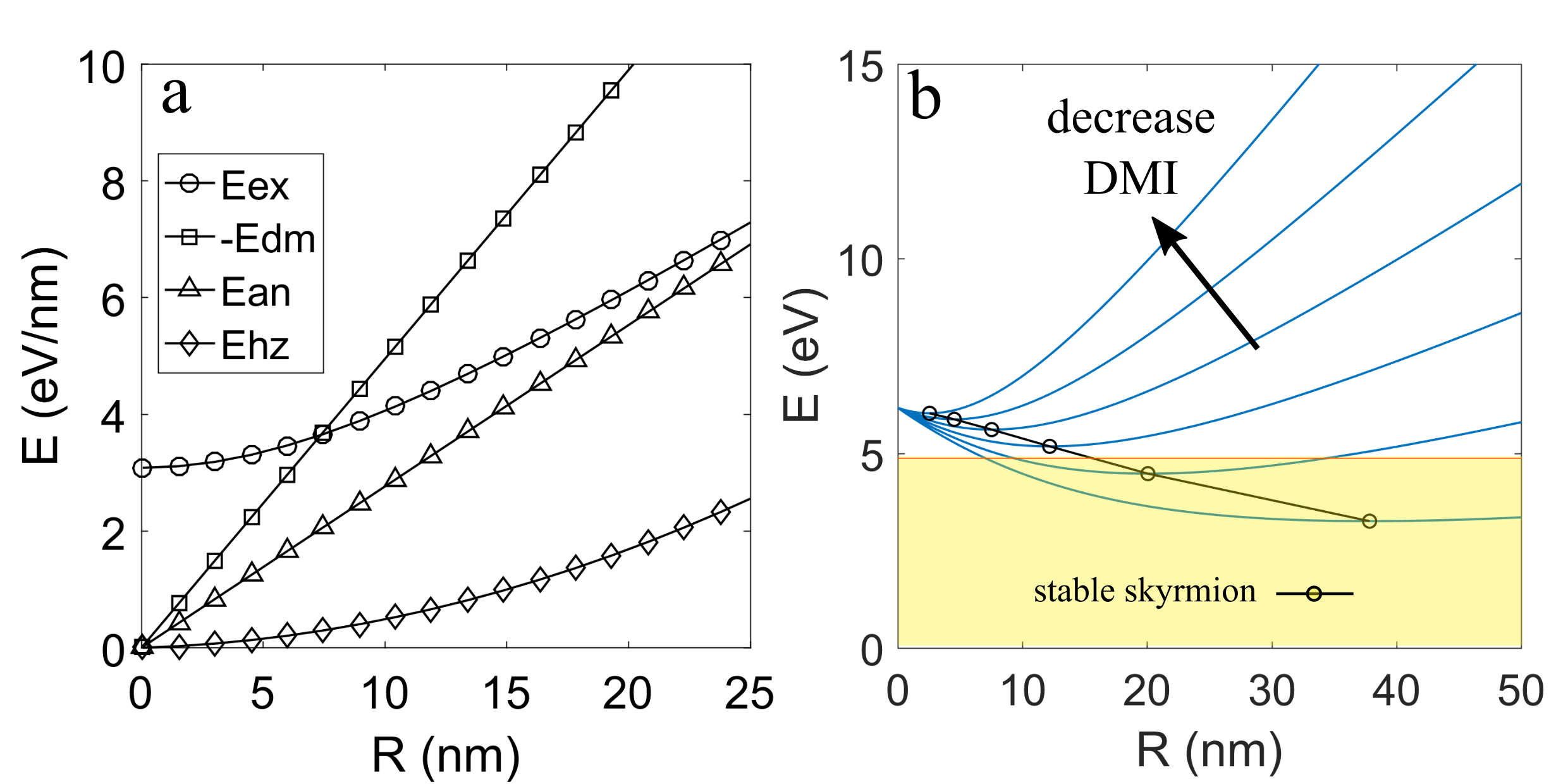}
\caption{(a) Individual skyrmion energy terms as a function of skyrmion Radius. Here the negative of DMI energy is plotted to allow easy comparison with other terms. The skyrmion wall width $\Delta$ is fixed in this plot. (b) Total energy as a function of skyrmion radius for different DMI with parameters $M_s=1e6\,\mathrm{A/m},\,K=3e5\,\mathrm{J/m^3},\,D\in[1.8, 3]\,\mathrm{mJ/m^2},\,t=2\,\mathrm{nm}$. The energy minimum marked by circles are skyrmion states. The energy of the ferromagnetic phase is chosen as the reference $E=0$. The top of the energy barrier between the skyrmion phase $R\rightarrow 0$ and the ferromagnetic phase is approximately $27.3At$\cite{buttner2018theory} the continuous limit. However, a discrete spin lattice can reduce the energy barrier top by $2\sim 4\,At$ depending on the lattice constant. The yellow region marks the stable skyrmion regime ($E_b\ge50k_BT$). Optimization of parameters can achieve the minimum stable radius below $5\,\mathrm{nm}$.}
\label{fig:skm_ene_dep}
\end{figure}

Under equilibrium conditions, the skyrmion size is determined by competing interactions from the exchange, Dzyaloshinskii-Moriya, anisotropy, stray field, and external magnetic field. In the continuous limit, the energy of an isolated skyrmion in a thin film can be written as:

\setlength{\arraycolsep}{0.0em}
\begin{eqnarray}
E={}&&t\iint \left\{A\left(\boldsymbol{\nabla}\mathbf{m}\right)^2-Km_z^2-\mu_0M_s\mathbf{m}\cdot\mathbf{H}_\mathrm{ext}\right. \nonumber \\
&&{}\left.\quad-\frac{1}{2}\mu_0M_s\mathbf{m}\cdot \mathbf{H}_\mathrm{d} + D\mathbf{m}\cdot[(\mathbf{\hat{z}}\times\vec{\nabla})\times\mathbf{m}]\right\}d^2\mathbf{r}
\label{eq:continuous_Hamil} 
\end{eqnarray}
with $t$ the film thickness, $A$ the exchange stiffness, $K$ the uniaxial anisotropy, $\mathbf{H}_\mathrm{ext}$ the external magnetic field and $\mathbf{H}_\mathrm{d}$ the stray field. The last term describes the interfacial DMI energy characterized by the coefficient $D$ and $\hat{\mathbf{z}}$ is the unit vector normal to the interface between the magnetic film and heavy metal top/bottom layer. At local equilibrium, the skyrmion state stabilizes at a local energy minimum. The solution to Eq.~\ref{eq:continuous_Hamil} can be obtained using standard micromagnetics packages (e.g. OOMMF\cite{donahue1999oommf}) or  approximate analytical forms\cite{buttner2018theory} that allow  fast and accurate evaluation of skyrmion texture and energy. Fig.~\ref{fig:skm_ene_dep}(a) depicts the dependence of different energy terms on the skyrmion radius. For simplicity of the discussion, the stray field energy is incorporated in the effective uniaxial anisotropy.

Before discussing the size and stability of skyrmions, it is important to distinguish between isolated skyrmions and skyrmion lattice. From a stability perspective, while skyrmion lattice can form a ground state, an isolated skyrmion, typically written with a current, is usually a meta-stable state with a finite energy barrier separating it from the homogeneous ferromagnetic phase. Significantly, isolated and lattice skyrmions have opposite size dependencies on material parameters such as DMI. Increasing just the DMI expands an isolated skyrmion until a critical DMI $D_c$, whereupon it transitions into a skyrmion lattice phase with each skyrmion cell shrinking with further increase in DMI. Such a non-monotonic size variation has indeed been observed numerically at 0K (skyrmion size diverges at $D_c$) \cite{rohart2013skyrmion,butenko2010stabilization} and finite temperature (smoother transition between two regimes) \cite{siemens2016minimal}. In this work, we limit our discussion with the isolated skyrmion regime due to their appeal in easy manipulation for potential memory/storage applications.

By treating the isolated skyrmion texture as a radial 1D $\pi$ domain wall, Rohart {\it et al} have derived a simple equation for the skyrmion radius\cite{rohart2013skyrmion} for isolated skyrmions:
\begin{equation}
R=\frac{\Delta_0}{\sqrt{2(1-D/D_c)}}
\label{eq:big_skm_R}
\end{equation}
where $\Delta_0=\sqrt{A/K}$ is the 1D domain wall width and $D_c=4\sqrt{AK}/\pi$ is the critical DMI value beyond which the skyrmion lattice or Neel stripe phase dominates. The assumption behind Eq.\ref{eq:big_skm_R} that $\Delta_0$ only depends on $A,K$ works well for large skyrmions $R\gg \Delta_0$ but not for smaller ones. In Appendix \ref{sec:analy_R_E}, we have fitted different energy integrals from the skyrmion Hamiltonian to come up with an analytical equation for skyrmion radius that also works for small skyrmions:
\begin{equation}
    R=\left(\frac{D}{D_c}\right)^3\frac{C_1\Delta_0}{\sqrt{1-C_2\left(\frac{D}{D_c}\right)^4}}
\label{eq:big_skm_R2}
\end{equation}
where $C_1\approx 2.233,\;C_2\approx 1.141$ are constants. We clearly see that the skyrmion size reduces as $D$ decreases (also see Fig.\ref{fig:skm_ene_dep}b). However, as the skyrmion size reduces, the energy barrier separating the skyrmion phase and the ferromagnetic phase ($E=0$) also diminishes, rendering the skyrmion unstable against thermal fluctuations at room temperature. There is thus a narrow window of phase space for a stable small isolated skyrmion.

\begin{figure}[ht]
\centering
\includegraphics[width=2.8in]{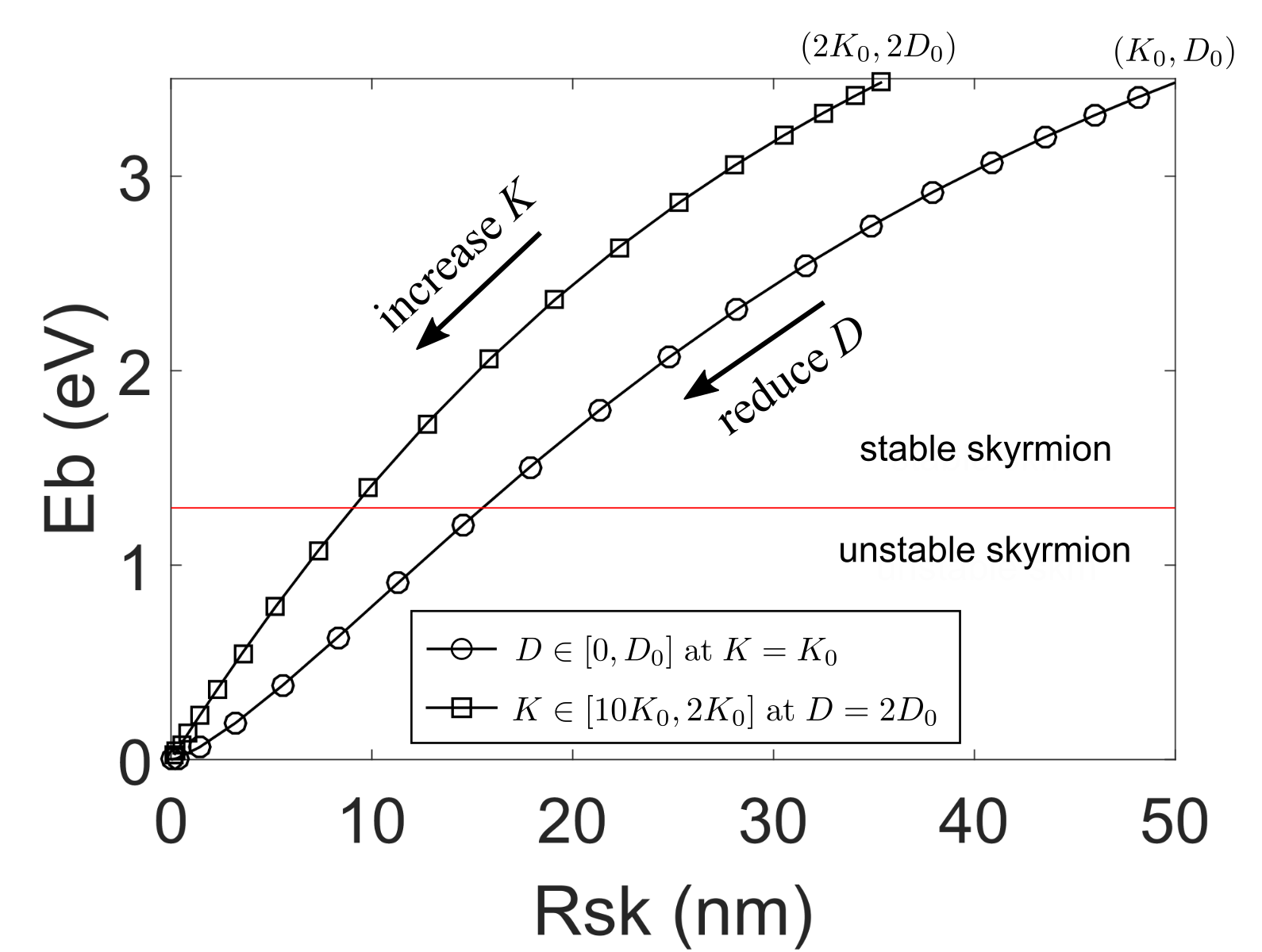}
\caption{The energy barrier separating the skyrmion phase and the ferromagnetic phase as a function of skyrmion radius. The skyrmion radius is tuned by varying $D$ or $K$. Reducing $D$ alone reduces the skyrmion size. However, the minimum radius also depends on other parameters described earlier. If the objective is to reach the smallest skyrmion at a fixed energy barrier for stability, this actually requires us to maximize $D$ while adjusting other parameters such as $K$.}
\label{fig:skm_ER}
\end{figure}

The required lifetime of a skyrmion is application specific. Here we consider a minimum barrier of $40k_BT-50k_BT$ to be stable. This is more of a conserved estimate because the actual lifetime of skyrmion depends on the attempt frequency, which is determined by various factors and can vary dramatically with external field\cite{von2018skyrmion}. This thermal stability constraint puts a lower limit on the skyrmion size. It is worth distinguishing the minimum skyrmion radius discussed here from the minimum radius due to the discrete nature of the atomic sites discussed in the literature \cite{siemens2016minimal}. The minimum skyrmion radius with adequate thermal stability can be quite large (10s of nanometers), while the minimum radius due to the discretization is very small ($<1\,\mathrm{nm}$). The thermal stability-constrained minimum radius depends on the material parameters as well as the film thickness. In Fig.~\ref{fig:skm_ER}, the energy barrier as a function of the skyrmion radius is plotted against different DMI and anisotropy constants $K$. It is important to point out that tuning individual material parameters independently might not be the optimal solution and the optimization strategy can depend on the objective: e.g. achieving the smallest skyrmion versus achieving an optimal phase space under constraints such as limited $D$ and $K$ ranges for small skyrmions. These trade-offs become clearer in sec\ref{sec:phase_space} when we examine the small skyrmion space for inverse Heuslers.

\subsection{Current-driven skyrmion mobility}
Skyrmions can be driven by an in-plane current through the heavy metal layer via spin-orbit torque, as has been demonstrated in multiple experiments\cite{litzius2017skyrmion,jiang2017direct,woo2016observation}. In the high speed regime, the skyrmion mobility (Eq.\ref{eq:mobility}) is determined by the skyrmion size, the saturation magnetization, and the skyrmion winding number. Since reducing the skyrmion size is our main objective here, lowering the saturation magnetization and the winding number is essential in improving the skyrmion mobility. For ferromagnets, the skyrmion winding number is integer 1. Ferrimagnets or antiferromagnets consist of two or more spin sublattices that oppose each other and the winding number averaged over all sublattices is less than one $\langle N\rangle \in [0,1]$. The mobility of the coupled skyrmions in a ferrimagnet can be derived from the Thiele equation\cite{buttner2018theory, thiele1973steady}: 

\begin{equation}
    \mu=\frac {\pi\gamma\hbar}{2e}\:\frac{I}{\sqrt{(4\pi) ^2 \langle N\rangle^2+\alpha^2{\mathcal{D}_{xx}^2} }}\frac{\Sigma_i\Theta_{\mathrm{sh,i}}}{\Sigma_i t_i\;M_{s_{i}}}
\label{eq:mobility}
\end{equation}
with $\gamma$ the gyromagnetic ratio, $t$ the thickness, and $e$ the electron charge. $\Theta_\mathrm{sh, i}$ is the material specific Spin Hall angle for sublattice $i$. $\mathcal{D}_{xx}=\int(\frac{\partial \mathbf{m}}{\partial x})^2 d^2r$,\;$I=\int(\sin{\theta}\cos{\theta}+r\frac{d\theta}{dr})dr$ where $\theta$ is the polar angle of the local magnetic moment. $\langle N\rangle=\Sigma_i N_it_iM_{s_{i}}/\Sigma_i t_iM_{s_{i}}$. The summation is over spin sublattices and $N_i=\pm 1$ is the winding number of skyrmion within each sublattice. $\overline{M}_s=\sum_it_iM_{s_i}$ is the net magnetization. In a synthetic antiferromagnetic structure, the top and bottom layers each host a skyrmion but with mirrored spin texture\cite{zhang2016thermally,zhang2016magnetic}. In those ideas, maintaining strong interlayer coupling and being able to drive both layers through a current can be challenge. An alternative is to use natural ferrimagnet or antiferromagnet. When a ferrimagnet is close to its magnetization compensation point $\langle N \rangle \rightarrow 1$, a low magnetic damping $\alpha$ also becomes increasingly important in boosting the mobility of the skyrmion. However, When the skyrmion is driven by the spin-orbit torque, the skyrmion Hall angle is determined by $\arctan\frac{-4\pi\langle N\rangle}{\alpha \mathcal{D}_{xx}}$. Reducing damping can increase the skyrmion Hall angle but in the limit $\alpha \ll \langle N \rangle$, the skyrmion moves in an orthogonal direction as the current, which could use a different circuitry for operation (e.g. othorgonal current injection). Other potential solutions to address the skyrmion Hall effect are based on film edge engineering or material composition engineering have been proposed\cite{fook2015mitigation,kim2017self}. 

The above-mentioned conditions for hosting small and fast skyrmions inspired us to look for ferrimagnetics/antiferromagnets in the Heusler family. The selection criteria include ferrimagnetic order (exists in Inverse Heusler), low saturation magnetization, high Neel temperature, and potentially half-metallic bandstructures. Half-metals are known to have low magnetic damping due to the lack of electron states in one spin channel \cite{liu2009origin,kumar2017temperature}. In the next section, we present half-metallic and near half-metallic ferrimagnets from our previous high-throughput studies of inverse Heuslers \cite{ma2017computational}, along with detailed calculations of their magnetic properties.

\section{Half-metallic Inverse Heuslers}

\subsection{Crystal structure}
\begin{figure}[t]
        \includegraphics[width=3.25in]{{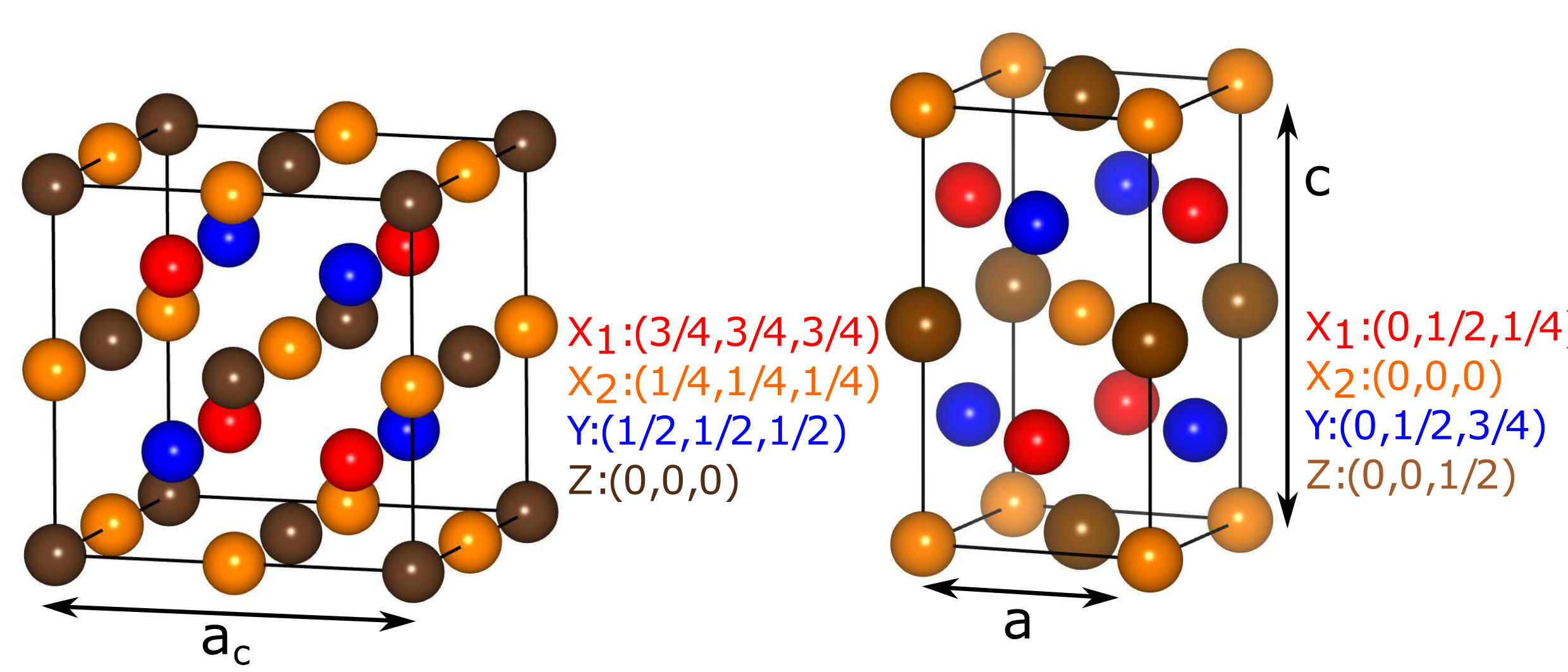}}
        \caption{Schematic representation of (a) cubic inverse-Heusler $\mathrm{XA}$ structure and (b) tetragonal inverse-Heusler structure. In the inverse-Heusler structure, $\mathrm{X_{1}}$ and $\mathrm{X_{2}}$ are the same transition metal element but they have different environments and magnetic moments.}
        \centering
        \label{fig:D2dXa}
\end{figure}

The inverse-Heusler compound, $\mathrm{X_{2}YZ}$, crystallizes in the face-centered cubic $\mathrm{XA}$ structure with four formula units per cubic unit cell. Its space group is no. 216, $F\bar{4}3m$. Its structure can be viewed as four interpenetrating FCC sublattices, occupied by the $\mathrm{X}$, $\mathrm{Y}$ and $\mathrm{Z}$ elements, respectively. The $\mathrm{Z}$ and $\mathrm{Y}$ elements are located at the (0,\, 0,\, 0) and $\left(\frac{1}{4},\, \frac{1}{4},\, \frac{1}{4}\right)$ respectively in Wyckoff coordinates, while the $X$ elements are at $\left(\frac{3}{4},\, \frac{3}{4},\, \frac{3}{4}\right)$ and $\left(\frac{1}{2},\, \frac{1}{2},\, \frac{1}{2}\right)$, resulting in two different rock-salt structures, $\mathrm{X_1Y}$ and $\mathrm{X_2Z}$ as shown in Fig. \ref{fig:D2dXa}(a). We use $\mathrm{X_{1}}$ and $\mathrm{X_{2}}$ to distinguish these two $\mathrm{X}$ atoms sitting at the two nonequivalent sites in the $\mathrm{XA}$ structure. The tetragonal inverse-Heusler structure can be obtained by stretching or compressing the parent cubic structure along the $z$ axis. The tetragonal unit cell shown in Fig. \ref{fig:D2dXa}(b) is rotated $45^\circ$ around the $z$ axis relative to the parent cubic structure. We can define the tetragonality as $t=c/a$. The lattice constant $a_{c}$  of the cubic structure can be obtained from $a$ as $a_{c} = \sqrt{2}a$.

\subsection{Neel temperature and Gilbert damping}
Fundamental electronic and magnetic properties of the Inverse Heusler familiy are investigated using the Vienna \textit{Ab Initio} Simulation Package (VASP) \cite{VASP1,VASP2} in our previous work\cite{ma2017computational}. Here we extend the investigations to include the exchange coupling and Gilbert damping via the framework of the Korringa-Kohn-Rostoker Green’s function formalism, as implemented in the Munich spin-polarized relativistic Korringa-Kohn-Rostoker (SPR-KKR) package \cite{EbertSPRKKRDamping1}. Relativistic effects were taken into account by solving the Dirac equation for the electronic states, and the atomic sphere approximation (ASA) was employed for the shape of potentials. An angular momentum cutoff of $l_{max} = 3$ (corresponding to $f$-wave symmetry) was used in the multiple-scattering expansion. A k-point grid consisting of $\sim$1000 points in the irreducible Brillouin zone was employed in the self-consistent calculation, while a substantially denser grid of $\sim$300000 points was employed for the Gilbert damping calculation. To achieve convergence, we used the  BROYDEN2 algorithm with the exchange-correlation potential of Vosko-Wilk-Nusair (VWN) \cite{vosko1980accurate}.

The atomistic exchange coupling $J_{ij}$ based on the classical Heisenberg model are obtained from the KKR method using  the Lichtenstein formula\cite{LIECHTENSTEIN_Jex}:
\begin{equation}\label{eq:2.50}
H_\mathrm{ex}=-\sum_{i,j}J_{ij}\mathbf{s}_{i}\cdot\mathbf{s}_{j}
\end{equation}
where $\mathbf{s}_i$ and $\mathbf{s}_j$ are unit vectors of the local magnetic moments on atomic sites $i$ and $j$. The calculated exchange couplings are site- and distance-dependent. An appropriate truncation of the cluster radius around each atomic site was chosen to be $4a$ ($a$ is the lattice spacing) to assure convergence of the Neel temperature, which was estimated through the mean field approximation (see Appendix~\ref{app:Neel temperature} for details). The Gilbert damping constant is also calculated with the Green's function method (see Appendix~\ref{app:damping}).

\begin{table*}[htbp]
\centering
        \caption{Structural and magnetic properties of Heusler compounds with low hull distance. Successive columns present: composition, calculated lattice constant, $a_\mathrm{Cal}$, experiment lattice constant, $a_\mathrm{Exp}$, saturation magnetization, $M_{S}$, formation energy $\Delta E_f$, distance from the convex hull $\Delta E_{\rm HD}$, experiment Neel temperature, $T_{N}(\mathrm{Exp})$, calculated Neel temperature $T_{N}(\mathrm{Cal})$, calculated Gilbert damping at room temperature, $\alpha$, and electronic ground state (Electronic ground state: M = nonmagnetic metal, HM = half-metal, NHM = Near half-metal). The last two columns show their potential tetragonal phase structure and energy difference from their cubic phase $\Delta E=E_\mathrm{cubic}-E_\mathrm{tetragonal}$.}
         \begin{tabular}{|c|c|c|c|c|c|c|c|c|c|c|c|c|c|c|}
    \toprule
              & \multicolumn{8}{c|}{Cubic phase}  & \multicolumn{2}{c|}{Tetragonal phase}   \\ 
    \hline
      $\mathrm{XYZ}$   & $a_\mathrm{Cal}$ & $a_\mathrm{Exp}$  & $M_{S}$ & $\Delta E_{\rm HD}$ & $T_{N}(\mathrm{Exp})$ &$T_{N}(\mathrm{Cal})$ & $\alpha$   &Electronic & $a, c$ & $\Delta E$  \\
      & \multicolumn{2}{c|}{$\mathrm{(\AA)}$}  & (emu/cc) & (eV/atom) &   \multicolumn{2}{c|}{(K)} & ($10^{-3}$)   &  ground state  & $\mathrm{(\AA)}$ & (eV/atom)    \\
    \hline
    \midrule
    Mn$_{2}$CoAl    & 5.735  &5.798\cite{PhysRevLett.110.100401}   & 393.5  & 0.036 & 720\cite{PhysRevLett.110.100401} & 845 & 4.04 &   HM & 3.76, 6.68 & -0.05\\
    Mn$_{2}$CoGa    & 5.76  &5.86\cite{PhysRevB.77.014424}   &  389.1  & 0 & 740\cite{UMETSU2015890} & 770 & 2.18 &   NHM & 3.71, 7.13 & -0.0103 \\
    Mn$_{2}$CoSi    & 5.621  &    &  627.3  & 0.018 &  & 460 & 3.01 &   HM & & \\
    Mn$_{2}$CoGe    & 5.75  & 5.80\cite{PhysRevB.77.014424}   & 590.6   & 0.03 &  & 471& 4.97 &   HM & 3.75, 6.84 &  0.0144\\
   Mn$_{2}$FeAl    & 5.75  &    & 195.3   & 0.008 &  & 380 & 8.14&   NHM & 3.67, 7.28 & 0.0026\\
    Mn$_{2}$FeGa    & 5.79  &  &  198.5  & 0.018 &  & 496 & 7.37 &   NHM & 3.68, 7.29 & 0.0331 \\
    Mn$_{2}$FeSi    & 5.60  &    &  424 & 0    &  & 71 &3.98 &   NHM & 3.56, 7.26 & -0.071 \\
    Mn$_{2}$FeGe    & 5.72  &    & 399.1   & 0    &  & 210 & 3.04& NHM & 3.62, 7.45 & -0.0164\\
   Mn$_{2}$CuAl   & 5.89  &   & 33.4  & 0.042 &  & 1145 & 1.84 &      Metal & & \\
    Mn$_{2}$CuGa    & 5.937  &   & 57.86  &  0.0208 &  & 1242 & 1.59 &      Metal & & \\
    MnCrAs    & 5.51  &  &0     & 0.083    &  & 985 & 1.23 &  HM & & \\
    \bottomrule
    \end{tabular}%

        \label{tab:Inverse_Heusler_HalfMetal}%
\end{table*}%

\subsection{Ferrimagnetic half-metals/near half-metals}
Our previous studies on Inverse Heusler and Half Heusler compounds have predicted a series of experimentally feasible half-metals and near-half-metals, determined by the short distance to their respective convex hull of stable phases (less than 0.052 eV/atom for Inverse Heusler, less than 0.1 eV/atom for Half Heusler)\cite{ma2017computational, PhysRevB.95.024411}. In Table \ref{tab:Inverse_Heusler_HalfMetal}, we selected 10 Inverse Heusler candidates with hull distance less than 0.052 eV/atom and 1 Half Heusler candidate less than 0.1 eV/atom. All of those compounds are ferrimagnets (with MnCrAs being antiferromagnet). Mn$_{2}$CoAl, Mn$_{2}$CoSi, Mn$_{2}$CoGe, and MnCrAs are predicted half-metals. Mn$_{2}$CoAl, Mn$_{2}$CoGa, Mn$_{2}$CuAl, Mn$_{2}$CuGa and MnCrAs have very high Neel temperature and two of those compounds have been confirmed with high Neel temperature in experiments\cite{PhysRevLett.110.100401,UMETSU2015890}. Our mean field calculations seem to overestimate the Neel temperature compared to  experimental values. 

We calculated the Gilbert damping at room temperature. Although not all compounds listed here are half-metals, the proximity of the Fermi energy to the energy gap in the minority spin channel limits the interband spin-flip scattering at the room temperature. Besides, the intraband spin-flip contributions are limited at high temperature. The limited spin-flip scattering results in a small damping coefficient  $10^{-4}\sim10^{-3}$. Similar low damping simulations and experimental results have been verified for the full-Heusler half-metal Co$_{2}$MnSi\cite{PhysRevB.93.094410} and $\mathrm{Co_2MnGe}$\cite{shaw2018magnetic}. 

We also calculated the tetragonal distorted structure for the 10 inverse-Heusler candidates, in order to look for intrinsic crystal anisotropy. Seven of them have local energy minima in the tetragonal phase. The energy differences between the cubic and the tetragonal phases are listed in table \ref{tab:Inverse_Heusler_HalfMetal}. Positive $\Delta E$ indicates the tetragonal phase is the global energy minima. It is worth mentioning that $\mathrm{Mn_2CoGa}$ has enhanced Neel temperature ($770\,\mathrm{K}\rightarrow975\,\mathrm{K(tetragonal)}$) and a much reduced saturation magnetization ($389.1\,\mathrm{emu/cc}\rightarrow27.6\,\mathrm{emu/cc(tetragonal)}$) in the tetragonal phase compared to its cubic phase, which makes it the most promising candidate among those in the table. For the cubic phase, an interface-induced anisotropy (or external field) is needed to achieve perpendicular magnetization.

\subsection{Phase space for small stable skyrmions}
\label{sec:phase_space}

\begin{figure*}[th]
\centering
\includegraphics[width=\textwidth]{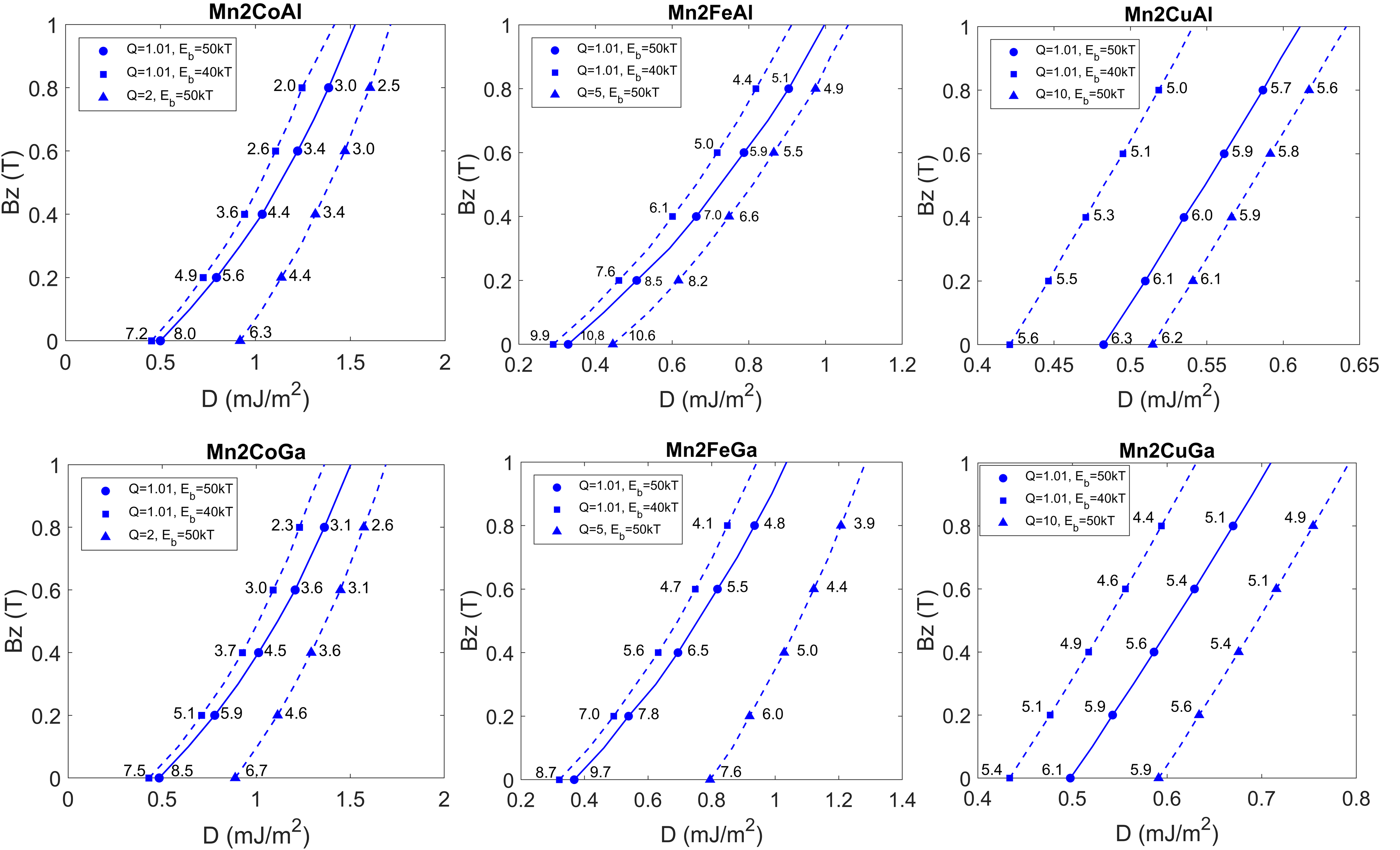}
\caption{Smallest stable skyrmion boundary for inverse Heuslers. The lines indicate the smallest skyrmions with an energy barrier $E_b=50k_BT$ or $E_b=40k_BT$ between the ferromagnetic state and skyrmion state. The scatter plot samples the skyrmion size along the boundary. The film thickness is assumed $5\,\mathrm{nm}$ in all calculations. $Q=2K/\mu_0M_s^2$ defines the effective anisotropy compared to the demagnetization and $Q>1$ is needed for a perpendicular system. To show the effect of anisotropy on skyrmion size and stability, different $Q$ are chosen for different materials due to varying saturation magnetization.}
\label{fig:phase_space}
\end{figure*}

We have down selected six compounds with relatively high Neel temperature and low saturation magnetization. In Fig.~\ref{fig:phase_space}, the smallest skyrmion given a fixed energy barrier $E_b$ is plotted against different external fields and DMIs. Here we used an effective ferromagnetic model with an average exchange stiffness extrapolated from their Neel temperatures. This approximation captures the energetics of the equilibrium skyrmion at 0K. The internal antiferromagnetic structure becomes important when spin dynamics is involved and the continuous model is no longer valid. The contour separates the stable skyrmion phase (for a given energy barrier criterion) and the unstable skyrmion or ferromagnetic phase. The effects of different energy barriers (i.e. skyrmion lifetimes) and uniaxial anisotropy on the skyrmion size are also presented. It is also worth mentioning that even though the skyrmions are simulated within a continuous model (OOMMF), a small correction ($\delta E=-2At$) of the top of the barrier between the skyrmion phase and the ferromagnetic phase needs to come in to account for the discrete nature of the atomic sites, where the continuous model is no longer valid near $R\rightarrow 0$. 

In Fig.\ref{fig:phase_space}, for a fixed barrier $E_b$ and anisotropy $Q$, one common observation is that the smallest skyrmion always exists at larger $B_z$ and DMI; increasing anisotropy helps reduce the skyrmion size but also requires a larger DMI. The need for larger DMI to achieve smaller skyrmion {\it{for a given energy barrier}} can be understood by revisiting the energy dependencies on radius in Fig.~\ref{fig:skm_ene_dep}. Due to DMI being the only term with a negative slope with respect to $R$, the energy barrier is approximately associated with DMI while the skyrmion radius with $dE\approx (D-K\Delta)R$. For a given $dE$, a larger DMI would allow smaller skyrmions if $B_z$ or $K$ are also increased concomitantly to push the energy minimum (i.e., skyrmion state) as close to $R=0$ as possible. Otherwise for a fixed $K-B_z$ pair, a larger DMI would only result in a bigger skyrmion, as discussed in section \ref{sec:analy_R_E}. Another way to improve the skyrmion stability is to increase the film thickness, which scales all energies including the energy barrier. However, considering that DMI is an interfacial effect which decays rapidly away from the interface, it would be hard to maintain decent DMI in thicker films or to apply appreciable interfacial spin-orbit torque to drive the skyrmions. A potential solution is to use a multilayered structure to enhance the effective DMI.

\section{Summary}
In this manuscript, we have discussed general directions on how to achieve small and fast skyrmions, and offered simple analytical results for navigating the multi-dimensional parameter space for the optimization problem. Based on this knowledge, we have scanned through the Heusler database and identified several promising candidates within the Inverse Heusler (and one from the Half Heusler) family with desirable properties, i.e. low $M_s$ ferrimagnet, low damping and high Neel temperature for hosting small skyrmions. Based on first-principles calculations, we have investigated the equilibrium skyrmion phase space for six candidates. A careful balance between DMI and anisotropy (or external field) is necessary to achieve the smallest skyrmion size while staying within practical requirements for the external factors such as magnetic field or DMI. Increasing film thickness helps stablize the skyrmions, but maintaining adequate DMI and spin torque in these thicker films will need further consideration.


%

\appendices
\section{Equilibrium skyrmion size and energy}
\label{sec:analy_R_E}
In the small skyrmion regime ($R\sim\Delta$), we can fit each energy term with a simple function. To obtain an analytical result, we ignore the Zeeman energy from the external magnetic field (which can be fitted with a quadratic equation): 
\begin{eqnarray}
\nonumber E_\mathrm{ex} & \approx & tA C_{e1} \sqrt{C_{e2}^2+R^2/\Delta^2}\\
\nonumber E_\mathrm{DMI} &  \approx  &  -tD C_{d} R \\
\nonumber E_\mathrm{an} &  \approx &  tK C_{a} R\Delta \\
\end{eqnarray}
where $C_{e1},C_{e2},C_d,C_a$ are fitted dimensionless constants and their values are in Table \ref{tab:const}.

\begin{table}[ht]
\renewcommand{\arraystretch}{1.3}
\caption{Fitted constants}
\label{table_example}
\centering
\begin{tabular}{|c|c|c|c|c|}
\hline
const. & $C_{e1}$ & $C_{e2}$ & $C_d$ & $C_a$ \\
\hline
value & 11.92 & 2.07 & 19.80 & 12.48 \\
\hline
\end{tabular}
\label{tab:const}
\end{table}
Solving the equilibrium skyrmion radius by making $\partial E/\partial \Delta=0$ and $\partial E/\partial R=0$ gives:
\begin{eqnarray}
\nonumber \Delta &=& \frac{C_d D}{2C_a K}  \\
u = \frac{R}{\sqrt{R^2+C_{e2}^2\Delta^2}} &=&   
\frac{C_d^2 D^2}{4 C_a C_{e1} K A} < 1 \\
\nonumber R_{sk} &=& \frac{C_{e2} \Delta }{\sqrt{ u^{-2} - 1 }} \\
\label{eq:small_skm_R}
\end{eqnarray}
for $u\ge1$, there is no local energy minimum and thus an isolated skyrmion is not stable. To compare Eq.\ref{eq:small_skm_R} with Eq.\ref{eq:big_skm_R} for large skyrmions, we can rewrite Eq.~\ref{eq:small_skm_R} in a similar form:
\begin{equation}
    R_{sk}=\left(\frac{D}{D_c}\right)^3\frac{C_1\Delta_0}{\sqrt{1-C_2\left(\frac{D}{D_c}\right)^4}}
\end{equation}
where $C_1\approx 2.233,\;C_2\approx 1.141$. $\Delta_0=\sqrt{A/K}$ is the 1D domain wall width and $D_c=4\sqrt{AK}/\pi$ is the critical DM value.

\section{Neel temperature}
\label{app:Neel temperature}

With the atomistic pair-wise exchange coupling, we can obtain the Neel temperature from the mean field approximation:

\begin{equation}
\frac{3}{2}k_{B}T_{C}^{MFA}=J_{0}=\sum_{j}J_{0j}
\end{equation} 

Since Heusler compounds have four sublattices, we have to solve the coupled equation:

\begin{equation}
\begin{aligned}
\frac{3}{2}k_{B}T_{C}^{MFA}\left\langle e^{\mu}\right\rangle &=\sum_{\nu}J_{0}^{\mu\nu}\left\langle e^{\nu}\right\rangle\\
J_{0}^{\mu\nu}&=\sum_{\boldsymbol{r}\neq0}J_{0\mathit{\boldsymbol{r}}}^{\mu\nu}
\end{aligned}
\label{coupledEquation}
\end{equation}

\noindent where $\left\langle e^{\mu}\right\rangle $ is the average $z$ component of the unit vector $e_{\boldsymbol{r}}^{\nu}$ in the direction of magnetic moment at site $(\nu,\boldsymbol{r})$. The coupled equation can be rewritten as eigenvalue equation:

\begin{equation}
\begin{aligned}
(\Theta-\boldsymbol{TI})\boldsymbol{E}&=0\\
\frac{3}{2}k_{B}\Theta_{\mu\nu}&=J_{0}^{\mu\nu}
\end{aligned}
\end{equation}

\noindent with a unit matrix $\boldsymbol{I}$ and the vector $\boldsymbol{E}^{\nu}=\left\langle e^{\nu}\right\rangle $. The Neel temperature can be calculated from the largest eigenvalue $J_{max}$ of the $\Theta$ matrix\cite{JmaxNeel1,anderson1963theory}. 

\section{Gilbert Damping}
\label{app:damping}

The Gilbert damping is calculated by the linear response formalism\cite{EbertSPRKKRDamping2}. The approach derives from the electronic structure represented by the Green function $G^{+}(E)$. $G^{+}(E)$ is determined by means of the multiple scattering formalism\cite{EbertSPRKKRDamping3}. The diagonal elements $\mu = x, y, z$ of the Gilbert damping tensor can be written as:

\begin{equation}
\alpha^{\mu\mu}=\frac{g}{\pi m_\mathrm{tot}}\sum_{j}\mathrm{Tr}\left\langle \mathcal{T}_{0}^{\mu}\tilde{\tau_{0j}}\mathcal{T}_{j}^{\mu}\tilde{\tau_{j0}}\right\rangle_{c}
\label{eq:damping}
\end{equation}

\noindent where the effective $g$ factor $g=2(1+m_\mathrm{orb}/m_\mathrm{spin})$ and the total magnetic moment $m_\mathrm{tot}=m_\mathrm{spin}+m_\mathrm{orb}$ is the sum of the spin and orbital moments $m_\mathrm{spin}$ and $m_\mathrm{orb}$ ascribed to a unit cell. Eq.~\ref{eq:damping} gives $a^{\mu\mu}$ for the atomic cell at lattice site 0 and implies a summation over contributions from all sites indexed by $j$, including $j=0$. $\tilde{\tau_{ij}}$ is given by the imaginary part of the multiple scattering operator that is evaluated at the Fermi energy $E_{F}$. $\mathcal{T}_{i}^{\mu}$ is represented by the matrix elements of the torque operator $\hat{\mathcal{T}}^{\mu}=\beta(\overrightarrow{\sigma}\times\hat{m}_{z})_{\mu}B_{xc}(\overrightarrow{r})$\cite{EbertSPRKKRDamping4}. The notation $\langle \cdots\rangle$ represents the configurational average including the vertex corrections\cite{PhysRevLett.107.066603} derived by Bulter\cite{PhysRevB.31.3260}. The configurational average accounts for finite temperature using the alloy analogy model within coherent potential approximation (CPA)\cite{PhysRevB.91.165132}.

\section*{Acknowledgment}
This work is funded by DARPA-Texitronics. Some of the tools were developed under NSF-SHF-1514219 and NSF-DMREF-1235230. We would like to thank Prof. Joseph Poon (UVa), Prof. Geoffery Beach (MIT), Prof. Andrew Kent (NYU) and Dr. Felix Buettner (MIT) for insightful discussions. We would also like to thank University of Virginia's High-Performance Computing service (Rivanna) for providing computing resources for this work.

\ifCLASSOPTIONcaptionsoff
  \newpage
\fi



%

\bibliographystyle{IEEEtran}
\bibliography{IEEEabrv,mybibfile}

%








\end{document}